# Handwriting Quality Analysis using Online-Offline Models

Yahia Hamdi[1,*] · Hanen Akouaydi[1] · Houcine Boubaker[1]
Adel M. Alimi[1]

**Abstract** This work is part of an innovative e-learning project allowing the development of an advanced digital educational tool that provides feedback during the process of learning handwriting for young school children (three to eight years old). In this paper, we describe a new method for children handwriting quality analysis. It automatically detects mistakes, gives real-time on-line feedback for children's writing, and helps teachers comprehend and evaluate children's writing skills. The proposed method adjudges five main criteria: shape, direction, stroke order, position respect to the reference lines, and kinematics of the trace. It analyzes the handwriting quality and automatically gives feedback based on the combination of three extracted models: Beta-Elliptic Model (BEM) using similarity detection (SD) and dissimilarity distance (DD) measure, Fourier Descriptor Model (FDM), and perceptive Convolutional Neural Network (CNN) with Support Vector Machine (SVM) comparison engine. The originality of our work lies partly in the system architecture which apprehends complementary dynamic, geometric, and visual representation of the examined handwritten scripts and in the efficient selected features adapted to various handwriting styles and multiple script languages such as Arabic, Latin, digits, and symbol drawing. The application offers two interactive interfaces respectively dedicated to learners, educators, experts or teachers and allows them to adapt it easily to the specificity of their disciples. The evaluation of our framework is enhanced by a database collected in Tunisia primary school with 400 children. Experimental results show the efficiency and robustness of our suggested framework that helps teachers and children by offering positive feedback throughout the handwriting learning process using tactile digital devices.

**Declarations**

**Compliance with Ethical Standards**

**Funding:** This study was funded by the Ministry of Higher Education and Scientific Research of Tunisia (grant number LR11ES4).

**Conflict of Interest**

The authors declare that they have no conflict of interest.

**Ethics approval**

This article does not contain any studies with human participants or animals performed by any of the authors.

* Corresponding author: Yahia Hamdi E-mail: yahia.hamdi.tn@ieee.org
Authors E-mail {yahia.hamdi.tn, hanen.akouaydi.tn, houcine-boubaker, adel.alimi}@ieee.org
[1] REGIM-Lab.: REsearch Groups in Intelligent Machines, University of Sfax, ENIS, BP 1173, Sfax, 3038, Tunisia.

## 1 Introduction

Nowadays, many studies have been developed in handwriting learning topic. It considered as one of the current challenges in multi-script handwriting analysis such as Arabic and Latin due to the variability of their characteristics and writing styles. After a comparative study, Jolly et al. [1] have shown that the number of children trained on digital devices has improved significantly compared to others trained on paper especially in fluency terms. In this context, this paper deals with the problem of handwriting quality analysis. It consists of developing an educational tool allowing to help teachers and children from primary schools in the learning process and handwriting evaluation.
More precisely, this study contributes to the innovative learning project that takes advantage of the digital devices in primary schools with multiple-sequence traces (Arabic and Latin letters, cursive Arabic word, symbol drawing, and digits) on two main aspects.

- It provides a simple interface that allows children to choose the target sequence and works independently with real-time and online feedback.
- It helps teachers assess children's writing skills and determine their difficulties.

The main objective of our project is to provide an advanced digital writing experience at school by using digital devices (tablets and tactile). The validation of the research project is founded on experiments conducted on primary school children's in Tunisia (see Fig.1) who were drawing the block of Arabic, Latin letters, Arabic words, and symbols as illustrated in and Fig.2.
The handwriting quality analysis depends generally on readability (shape), order, and direction aspects. Indeed, at the beginning of the learning stage, it is more important to have powerful constraints on order and direction aspects to obtain correct handwriting coordination. However, shape represents the main aspect to be evaluated during the learning process. The score given to the shape aspect decreases consecutively with symbol deformation. Contrary, scores related to direction and order aspects correspond to a binary decision. In fact, education experts do not forever agree on alone writing convention but think that children must follow well-defined rules to know how to write correctly any symbol.

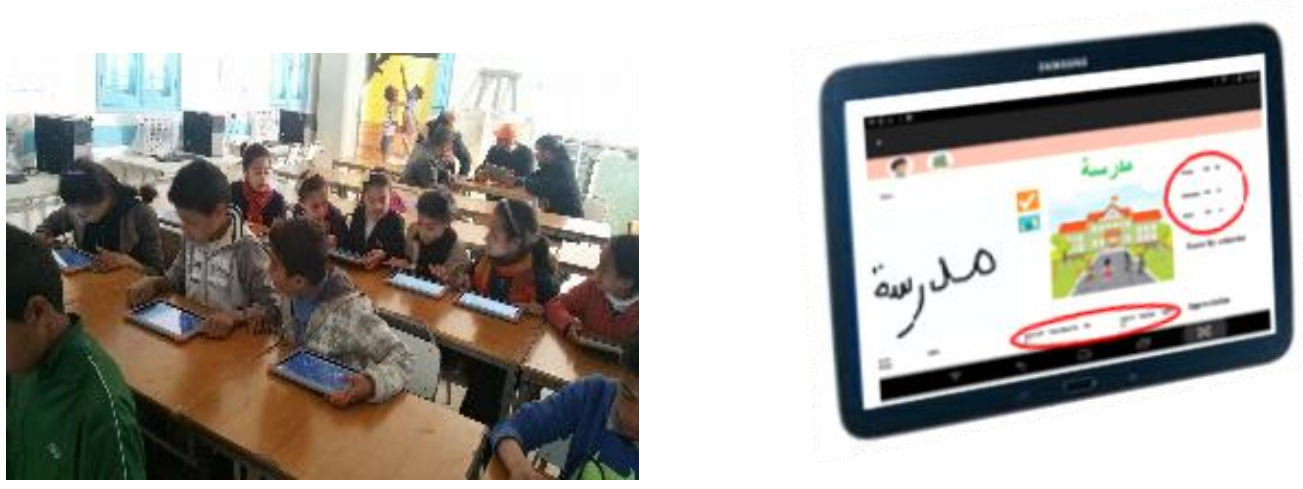

(a) (b)

**Fig. 1.** Experiment in-class of the proposed project (a), with tactile tablet devices about the analysis of supported scripts (b).

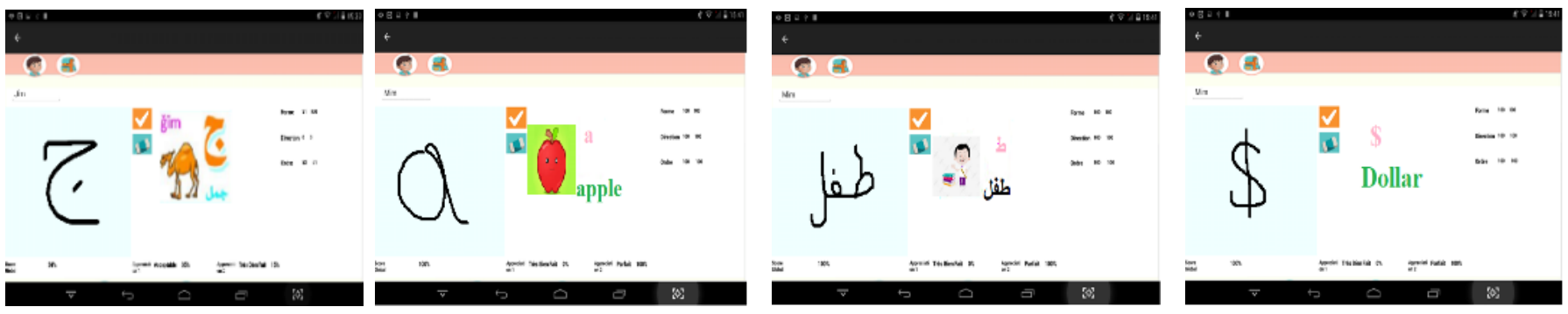

(a) Arabic Letter (b) Latin Letter (c) Arabic word (d) Symbol

**Fig. 2.** The online handwriting analysis framework that enables children to be autonomous during the learning and analysis tasks of letters (a, b), words (c), and symbols (d).

One of the most important problems of handwriting quality analysis is the selection of the relevant set of features allows characterizing such criteria. For instance, fuzzy histogram of direction and orientation, histogram of points are used by [2] to identify finely shape and direction of letter. Recently, deep learning technologies have proven highly successful in different topics of computer vision such as pattern recognition. CNN is one of the most widespread kinds of neural networks applied for action recognition [3], [4], handwriting recognition [5][6], etc. It is considered as a powerful feature extraction model that extracts automatically multiple low and high-features [7], [8]. This is the reason why, in this framework, we focus on CNN to extract efficient deep learning features allowing to characterize and analyze online children's handwriting letters shape with different writing style.

The main contributions of our work consist of developing a framework for multi-scripts handwriting quality analysis and gives online feedback regards to reference models. We have employed five main criteria (*shape, direction, order, kinematics, position respect to the reference lines*) with specific types of features (see section 5) that allow characterizing the children writing style: BEM, FDM, and deep learning-based CNN feature. The former is used for elementary stroke segmentation of the online handwriting trajectory and to apprehend among the dynamic features related to the motion and the geometry of the elementary stroke and their sequence time. The second analysis feature allows us to describe the overall geometry of the trace as a function of the undulations of its curvature function. Finally, CNN model is utilized for the perceptual capture of the post-drawing trace.

This paper is organized as follows. Section 2 depicts the state of the art of some works in handwriting quality analysis. In section 3, we introduce the proposed framework architecture and used criteria analysis. Section 4 summarizes the preprocessing techniques that are used in our work. The handwriting feature extraction models and comparison engines are described in section 5. The evaluation of our system and the experimental results are discussed in Section 6. Finally, section 7 provides a conclusion and future work.

## 2 Related work

Handwriting quality analysis is related to readability (*shape*) and kinematics in the literature [9]. Indeed, the handwriting readability corresponds to the shape of the letter, while, kinematics is focused on the writing process (eg, direction, order, fluidity) to be effective because writing is a fundamental skill needed to learn and use knowledge.

Descriptions of many criteria that represent the critical components of handwriting readability are discussed in [10], including slant; size; the degree of line straightness; spacing; shape (letter form); and the general merit of the writing. They deduce that computer analysis is more sensitive, accurate and reliable than subjective analysis, although they observe that present practical applications are still restricted.

There are generally two types of applications resulting from the handwriting analysis: education and medical systems. Guinet et al. [11] interested on kinematic aspect (i.e. duration, velocity, fluency) to distinguish handwriting pathologies. Jolly et al. [1] focused on handwriting velocity in order to indicate disorders of development coordination. Various features (e.g. pen lift duration, peak velocities and number of strokes per letter) are used by [12] to investigate handwriting kinematics of children on digital devices like tablets. It also shows that four kinematic domains: (automation and motor planning plan, velocity, spatial arrangement) are necessary for the best handwriting quality. Another educational system for Chinese characters is introduced by [13] which is identified by three types of errors such as stroke production errors (i.e. broken strokes, stroke reversal, succession of separate strokes), stroke relationship errors (e.g. length and position), stroke sequence errors (i.e. wrong sequencing of components, wrong stroke sequences in a component). Falk et al. [14] adopt five primitives (i.e. size, form, space, alignment, and legibility) to quantify children's writing skills. More generally, the kinematic aspect is used mainly by medical systems like [1], [11] by opposition to educational systems [12], [13] which is particularly interested in legibility features.

Recently, Simonnet et al. [2] proposed a multi-criteria approach for Latin handwriting quality analysis. In this work, the handwriting children are evaluated with regards to reference models using intra-class and inter-class scores. Indeed, a multi-criteria score describes the legibility (shape) and kinematic (order and direction) aspects for children, according to the teacher expectations. In Simonnet et al. [15] the authors introduced an explicit segmentation approach for handwritten cursive word evaluation. They start by identifying the primary segmentation hypotheses to reduce error propagation by adding a verification step through supervision. Next, they extract the letter hypotheses based scoring followed by word hypothesis extraction and evaluation by combining elastic matching and writing analysis scores. More recent, the authors [16] uses the fusion of multiple channels CNN networks for online children handwriting recognition. They convert the online signal into multiple image channels taking into account he dynamical information to improve the performance of a CNN.

## 3 Handwriting quality evaluation method

The present framework proposes an algorithm for analyzing the quality of children's handwriting. It consists of the adoption of multiple evaluation criteria and the application of powerful models to extract effective and complementary features vectors. Our goal is to help teachers evaluate handwriting and provide remedial feedback to children's schools during the learning process. As shown in Fig.3, the handwritten sequences are online signals captured with digital devices that have undergone a pre-processing step. Then, five main criteria are employed to evaluate the quality of handwriting based on the combination of three models of handwriting modeling: BEM, FDM, and CNN. Finally, a significant appreciation score is established for each criterion related to reference models (database of samples) based on SD-DD and SVM engines respectively.

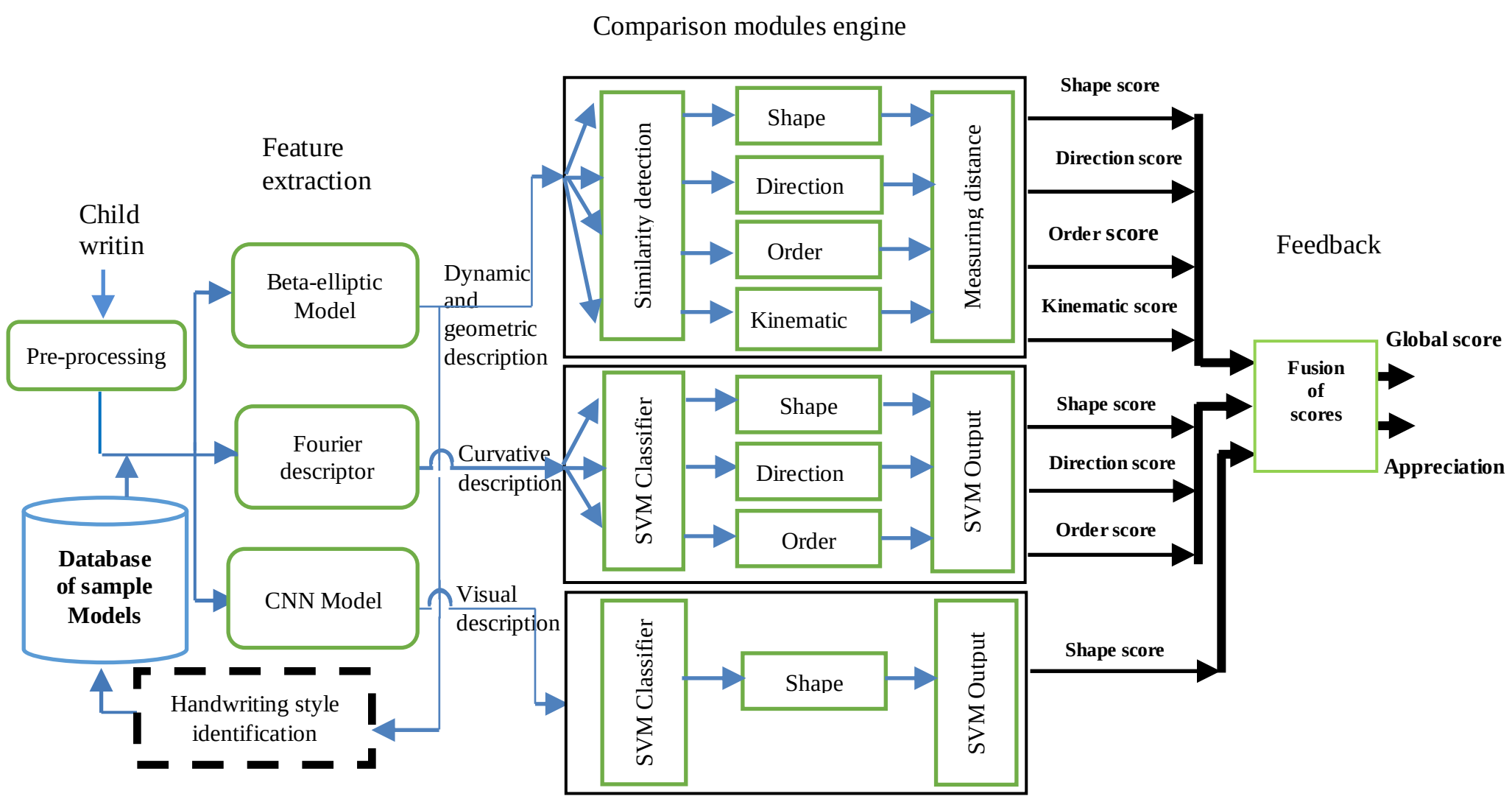

**Fig. 3.** Flowchart of the proposed framework.

### 3.1 The evaluation criteria

In this section, we briefly outline the list of evaluation criteria adopted by the proposed system for handwriting quality analysis.

#### 3.1.1 Graphical shape

The handwriting shape criterion is the main component of the handwriting quality assessment process. It corresponds to the fusion between the levels of graphical fidelity and readability of the produced writing taking into account the size and alignment on the guideline. Some geometric parameters such as the locations of a set of trajectory guide points that includes the segments endpoints and their intermediate points corresponding to velocity or curvature local extremums (see Fig.4.a and Fig.4.b for the correct and incorrect shape of the Arabic character 'ط' respectively). The global geometry of the trajectory segments and their curvature functions are also considered among the properties that characterize the shape criterion.

#### 3.1.2 Strokes order and direction

We suggest that the correct writing order of the sequence (letter, word, symbol) corresponds to the correct succession of the composed elementary strokes. Indeed, each script has its own properties distinguished from others. For instance, the Arabic language has several characters that contain dots (i.e: ب, ) or small marks (ط ,ئ) and other without dots (ع ,ح). We always start with the main corps before the diacritics. Fig.4. d presents an example of wrong order detected by our system which the segment 'ا' is written before the occlusion 'صـ'.

Also, the correct direction is identified using specifics features such as start and endpoint positions, and their trajectory tangents directions. (see Fig.4.c).

### 3.1.3 kinematics of the trace

The handwriting movement is considered as one of the most aspects of handwriting modeling systems that synthesis the neurophysiologic characteristics of the involved organs (muscles, hand, arm, finger joint, ...). A lot of research examine the kinematic of the handwriting trajectories and concentrate most of the information to be extracted in velocity profiles. According to this description, we believe that the concept of kinematics is very important to evaluate the dynamism of children 's writing. As shown in Fig.4.e, we notice a variation between regular and irregular writing kinematics, since this causes some diversity in the interpolation points of the trajectory.

| Handwritten segments | Trajectory sampling | Velocity profile | CS | CD | CO | CK | CP |
|---|---|---|---|---|---|---|---|
| 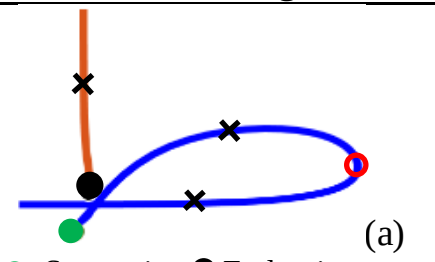  (a) | 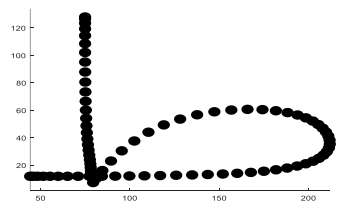 | 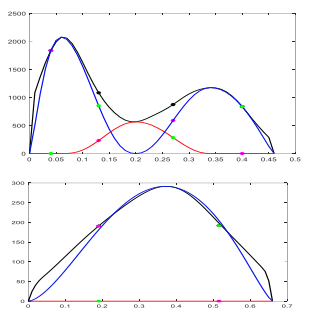 | ● | ● | ● | ● | ● |
| 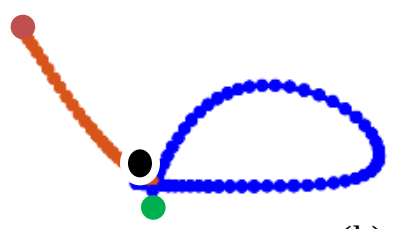 (b) | 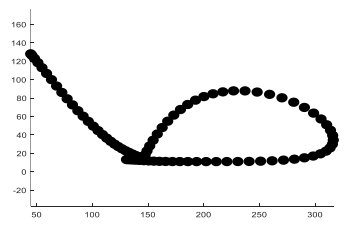 | 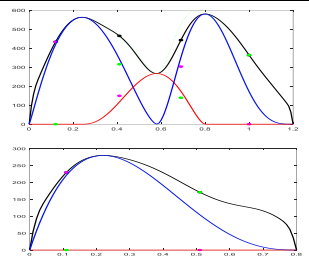 | ○ | ● | ● | ● | ● |
| 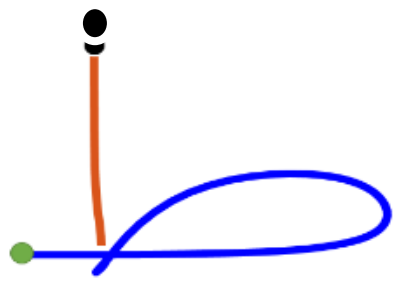 (c ) | 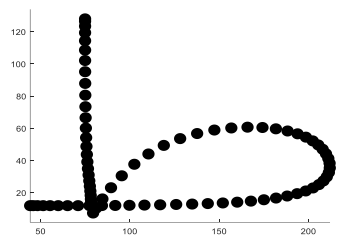 | 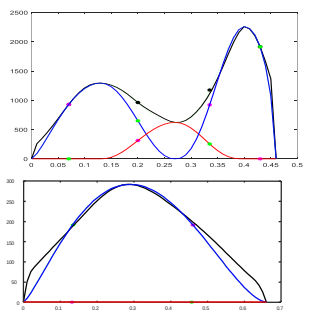 | ● | ○ | ● | ● | ● |
| 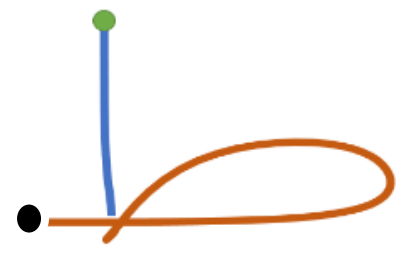 (d) | 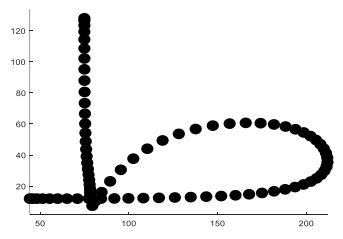 | 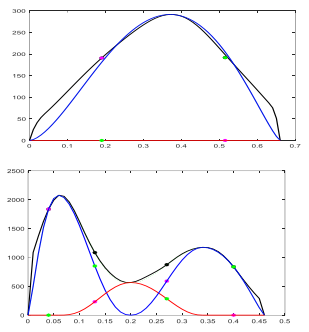 | ● | ● | ○ | ● | ● |
| 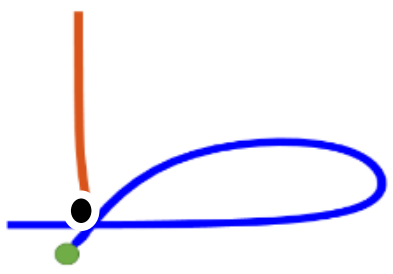 (e) | 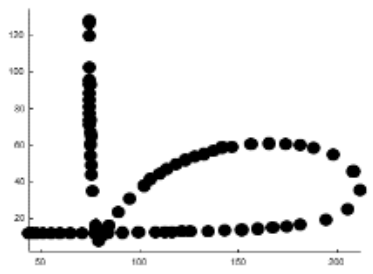 | 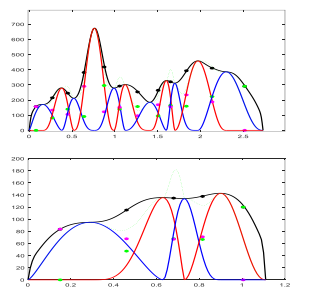 | ● | ● | ● | ○ | ● |
| 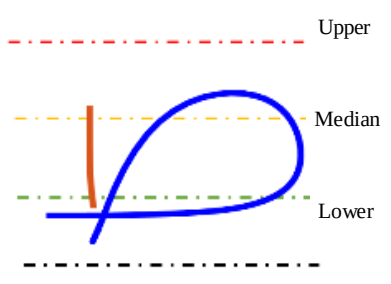  (f) | 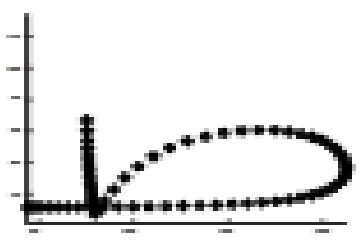 | 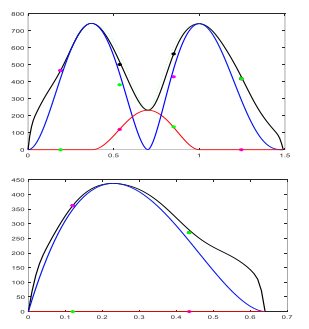 | ○ | ● | ● | ● | ○ |

CS: correct shape, CD: correct direction, CO: correct order, CK: correct kinematic, CP: correct position to respect reference line.

**Fig.4.** Presentation of the evaluation criteria: The first column represents the handwritten segments, while the second describes the temporal trajectory sampling. The third column shows the velocity profile of the correspondent trajectory sample. The last five columns depict respectively the correctness degree (black and white points for commonly correct and wrong criterion respectively) of the reported criteria.

### 3.1.4 Respect of the reference lines (trace position)

At the beginning of the school children's learning process, the baseline aspect is needed to learn and assess the competence of linear cursive writing. It is one of the most significant elements of handwriting analysis. It helps to determine how to deal with the combination of cognitive, social, and instinctual drives. Based on this criterion, the online signal can be decomposed in three main zones: Median, Upper, and Lower (see Fig.4.f) which allows the child to distinguish between the different forms of letters: those written on, below, and above the baseline.

## 3.2 Selection of the pertinent parameters

The evaluation of the quality of learners' writing tests depends on the five criteria presented above. The proportional progression of scores for all criteria during the learning process is not guaranteed since the learner often tends to focus on one or two criteria at the same time. On the other hand, the considered diversification of the evaluation criteria requires analogously a diversity of parametric models. Thus, the aim is to quantify the evaluation outcome of each separate criterion by computing its different associated parameters in order to deliver a final information report to the learner or the expert.

To ensure a strong consistency between the different analysis criteria, we applied the combination of three complementary models. In fact, we take advantage of BEM characteristics to apprehend the dynamic and graphical features of the elementary beta strokes (delimited by two successive extremums of velocity or curvature radius) composing the hand drawing trajectory. Further, the FDM and CNN models are used to describe respectively the variation of the trajectory curvature function and the final view obtained post-drawing.

However, to assess the compliance with the two criteria direction and order, we set aside the offline CNN model to keep only the BEM parameters as well as the FDM curvature model. Likewise, to evaluate the accordance with the kinematics criterion, we relied only on the dynamic features subset of BEM to compare the velocity profile of the test script to existing models recorded by an expert (voluntary teachers). The details of the different models employed are described in section 5.

# 4 Pre-processing

The handwriting trajectories are collected online via a digitizing device. It is characterized by a high variation which requires to apply geometric and denoise processing steps to minimize handwriting variabilities and reduce noise. Given the raw trajectory, the low-pass filtering Chebyshev type II with a cut-off frequency of fcut =10Hz is used to mitigate the effect of noise and errors due to temporal and spatial quantification introduced by the acquisition system as shown in Fig.5. The value of cut-off frequency results from a compromise between the conservation of the handwriting undulations produced by young writers and the elimination of those due to the pathologic tremor of other older writers [17].

The horizontal fulcrums level of character handwriting decomposes theirs drawing area in three zones namely upper, core and lower regions respectively. Consequently, procedure for normalizing the size of the handwriting is applied to adjust its height to a fixed value h = 128, while keeping the same ratio length/height [18], [19]. Both the preprocessing technique are used and tested for the supported scripts.

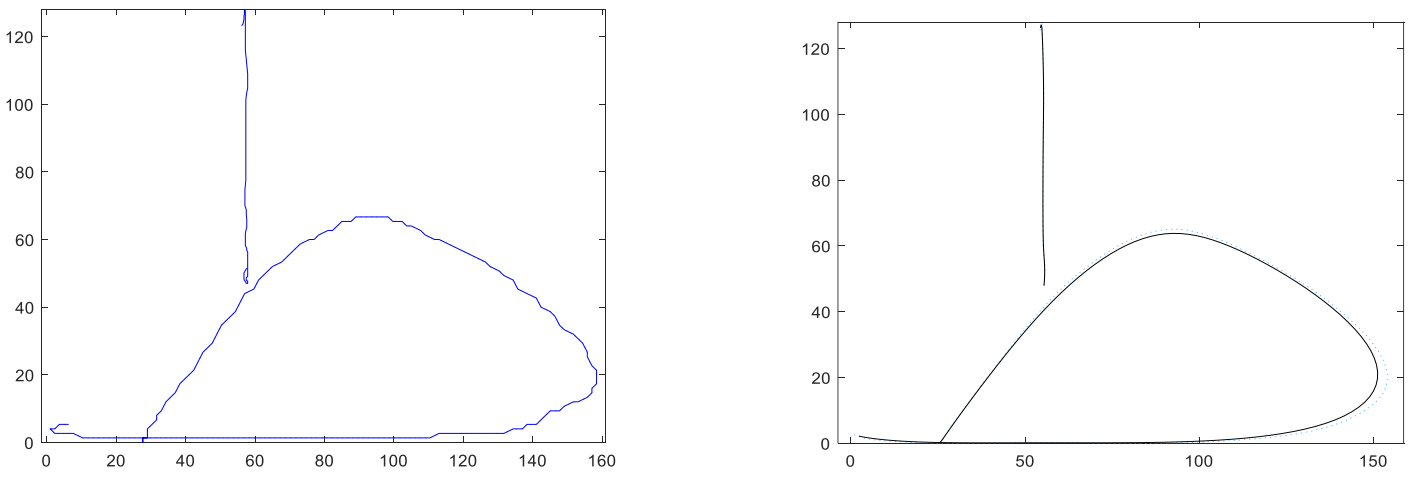

(a) (b)

**Fig.5.** (a) Acquired raw trajectory of the Arabic character 'ط' and (b) after Low-pass filtering and smoothing.

## 5 Features extraction and comparison engines

Our developed handwriting analysis system uses a new method based on the combination of handcrafted and deep learning feature extraction models coupled with different criteria assessment engines. These modules are described in turn in this section.

### 5.1 Features extraction

Three specifics types of features are introduced in our work: BEM, FDM, and CNN models. We explain in this sub-section the principal of each model.

#### 5.1.1 Beta-elliptic model (BEM)

The BEM derives from the kinematic Beta model with a juxtaposed analysis on the spatial profile. It considers a simple movement as the response to the neuromuscular system which is described by the sum of impulse signals [20], [21], the Beta function [22]. It consists of decomposing the online trajectory into elementary stroke based on the combination of the velocity profile (dynamic features) and static profile (geometric features) of online handwriting modeling. Further, BEM has proven highly successful in various areas of research on online handwriting like the study of the effect of age on hand movement kinematics [23], handwriting recognition [24], temporal order recovery [25], and writer identification [26]. To our knowledge, the BEM has not yet used in handwriting evaluation. In fact, our idea is to exploit the efficiency of the real-time description of handwriting movements provided by this model in our context.

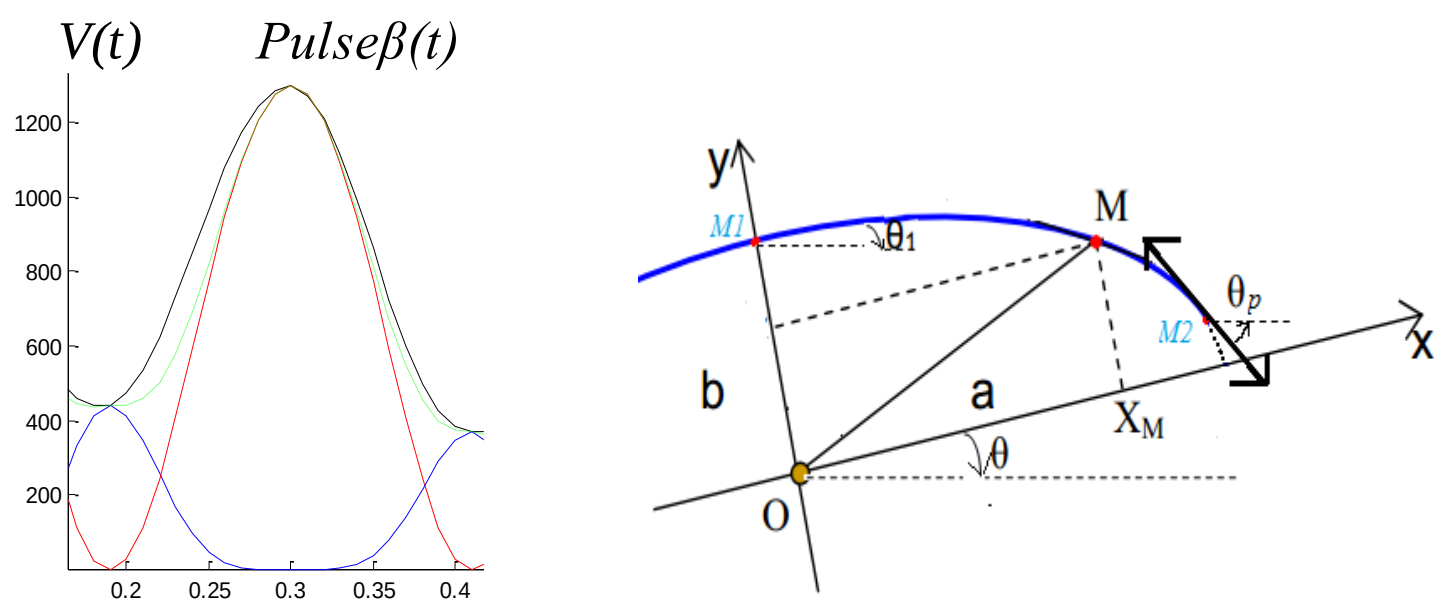

(a) Beta function (b) Elliptic arc

**Fig. 6** Online handwriting modeling using BEM.

**Velocity profile:**

In the dynamic profile, the curvilinear velocity curve $V\sigma(t_c)$ shows a signal that alternates between extremums (minima, maxima, and inflexion points) which delimit and define the number of trajectory strokes. It can be reconstructed by overlapping Beta signals where each stroke corresponds to the generation of one beta impulse (see Fig.6.a) represented by the following expression:

$$pulseB(t,q,p,t_0,t_1)=\begin{cases} k.\left(\frac{t-t_0}{(t_c-t_0)}\right)^p.\left(\frac{t_1-t}{(t_1-t_c)}\right)^q & if \quad t\in[t_0,t_1] \\ 0 \quad elsewhere \end{cases} \tag{1}$$

$$t_c=\frac{p*t_1+q*t_0}{p+q} \tag{2}$$

Where, $t_0$ and $t_1$ are respectively the starting and the ending times of the generated impulse which delimiting the correspondent trajectory stroke, $t_c$ is the instant when the beta function reaches its maximum value as depicted in Eq. 2, $K$ called impulse amplitude, $p$ and $q$ are intermediate shape parameters.

$$V_\sigma(t)=\sum_{i=1}^{n}V_i(t-t_{0i})\approx\sum_{i=1}^{n}pulse\beta i(K_i,t,q_i,p_i,t_{0i},t_{1i}) \tag{3}$$

As shown in Fig. 7.a) and c), the velocity profile of the handwritten trajectory of the Arabic character 'ك' and Latin word 'fleur' can be reconstructed by the overlapped beta signals as described in Eq. 3. Indeed, the dynamic features give the overall temporal properties of the neuromuscular networks implicated in the motion generation of the writing process.

**Geometric profile:**

In the space domain, each elementary beta stroke located between two successive extrema speed times can be modeled by an elliptic arc characterized by four geometric parameters: $a$, $b$, $\theta$, $\theta_p$ (see Fig.6.b). Where $a$ and $b$ represent respectively the half dimensions of the large and the small axes of the elliptic arc, $\theta$: the angle of the ellipse major axe inclination, and $\theta_p$ is the trajectory tangent inclination at the minimum velocity endpoint. These parameters reflect the geometric properties of the end effector (pen or finger) trace, dragged by the set of muscles and joints involved in handwriting.

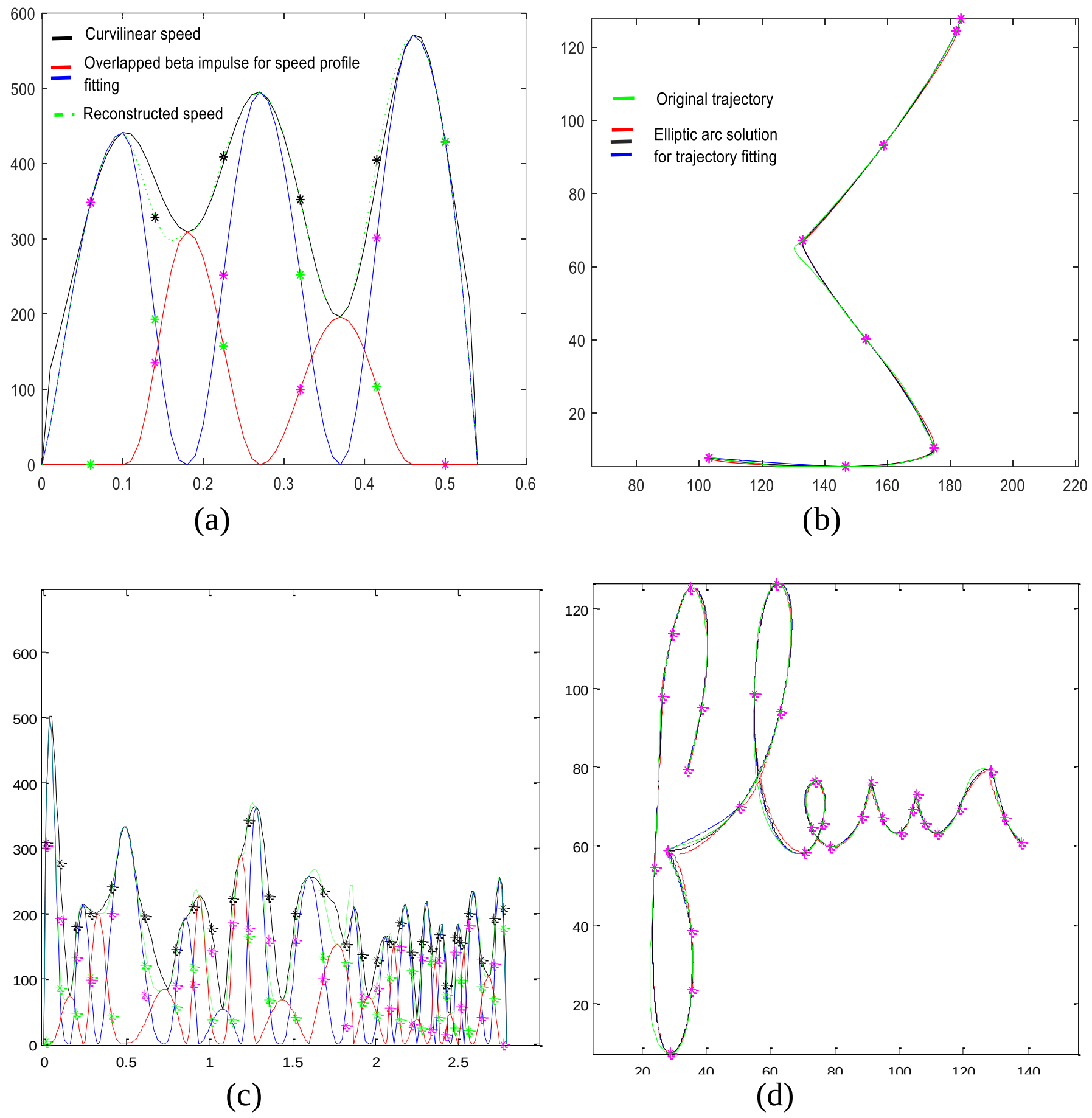

**Fig.7.** Online handwriting modeling of Arabic character 'ك' and Latin word 'fleur' with BEM. (a) and (c) describe the velocity profile, (b) and (d) represent the geometric profile.

### 5.1.2 Fourier descriptor model (FDM)

FDM is one of the most accurate tools for parametric modeling of the closed path which can be represented by a $2\pi$ periodic signature function [27]. To benefit from their powerful capacity of periodic function approximation in trajectories stroke modeling, we must transform the signatures corresponding to the stroke open trajectories into periodic functions. Each segmented trajectory is represented by an angular signature modeled by a periodic sinusoidal function. As shown in Fig. 8, the chosen function as a stroke trajectory signature, describe the variation of the inclination angle $\theta_i$ of the trajectory tangent at a point $M_i$ depending to its corresponding curvilinear abscissa:

$$\ell_i = \sum_{j=1}^{i} dL_j \quad for \quad i = 1, \ldots, 2n \tag{4}$$

Where, $dL_i$ represent the distance between the current point $M_i$ and its previous. Further, the Fourier descriptors coefficients $a_0$, $a_j$ and $b_j$ of the Fourier series that approximates the signature function $\theta_i = f(\ell_i)$ at the 8[th] harmonic are computed by Eqs. (5) and (6):

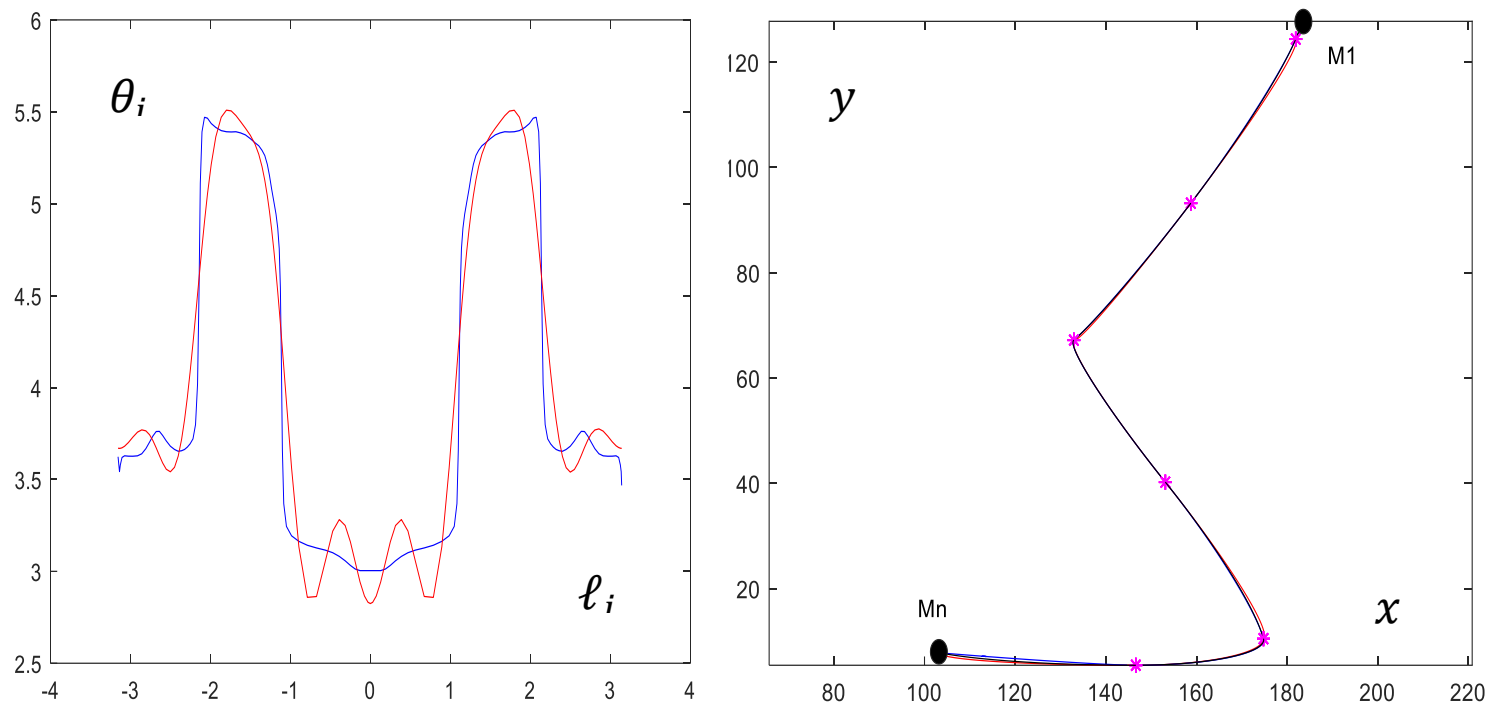

**Fig.8**. Signature functions approximated at the 8th harmonic and correspondent trajectory of Arabic character "ك".

$$a_0 = \frac{1}{2\pi} \cdot \sum_{i=1}^{2n} \theta_i \cdot dL_i \tag{5}$$

$$\begin{cases} a_j = \frac{1}{\pi} \cdot \sum_{i=1}^{2n} \theta_i \cdot \cos\left( j \cdot \frac{2 \cdot \pi \cdot \ell_i}{\ell 2n} \right) \cdot dL_i \\ b_j = \frac{1}{\pi} \cdot \sum_{i=1}^{2n} \theta_i \cdot \sin\left( j \cdot \frac{2 \cdot \pi \cdot \ell_i}{\ell 2n} \right) \cdot dL_i \end{cases} \tag{6}$$

To reconstruct the trajectory segment and the correspondent signature, we use the approximation function described by the following equation:

$$\theta_i = f\left(\ell_i\right) \approx a_0 + \sum_{j=1}^{k} \left[ a_j \cdot \cos\left( j \cdot \frac{2 \cdot \pi \cdot \ell_i}{\ell 2n} \right) + b_j \cdot \sin\left( j \cdot \frac{2 \cdot \pi \cdot \ell_i}{\ell 2n} \right) \right] \tag{7}$$

### 5.1.3 CNN shape model

In this sub-section, we briefly explain the use of CNN model in our context. Inspired by the recent successes of deep learning technologies in different topics, we have utilized CNN architecture in the off-line bitmaps reconstructed from the online trajectory. This later represents just the final layout of the hand drawing that skirt the chronologic style of its generation. The first experiment consists of selecting the best CNN feature extraction. Indeed, the choice of CNN architecture affects the performance of such system. The structure of CNN employed contains multiple convolution and max-pooling layers. We use also a recent dropout technique for selecting the best CNN feature extraction that characterizing shape model.

**Table 1. CNN configuration: k, s and p represent respectively the kernel size, stride and padding size.**

| Type | Configurations |
|---|---|
| Input | 32x32 |
| Convolution | #maps: 50, k: 3 × 3, s:1, p:1, |
| Convolution | #maps: 100, k: 3 × 3, s:1, p:1 |
| Convolution | #maps: 100, k: 3 × 3, s:1, p:1 |
| Max-Pooling | Window: 2 × 2, s: 2 |
| Convolution | #maps: 150, k: 3 × 3, s:1, p:1 |
| Convolution | #maps: 200, k: 3 × 3, s:1, p:1 |
| Convolution | #maps: 200, k: 3 × 3, s:1, p:1 |
| Max-Pooling | Window: 2 × 2, s: 2 |
| Convolution | #maps: 250, k: 3 × 3, s:1, p:1 |
| Convolution | #maps: 300, k: 3 × 3, s:1, p:1 |
| Convolution | #maps: 300, k: 3 × 3, s:1, p:1 |
| Max-Pooling | Window: 2 × 2, s: 2 |
| Convolution | #maps: 350, k: 3 × 3, s:1, p:1 |
| Convolution | #maps: 400, k: 3 × 3, s:1, p:1 |
| Convolution | #maps: 400, k: 3 × 3, s:1, p:1 |
| Max-Pooling | Window: 2 × 2, s: 2 |
| Full connection | #hidden units: 1600 , dropout: 0.2 |
| Full connection | #hidden units: 200 , dropout: 0.2 |
| Softmax | #units: 7357 |

After tried several settings, we chose CNN with 15 layers similar to Zhang et al. [28], because it is the most efficient architecture which generates a high accuracy for sequence recognition. The input layer is of size 32x32 of gray-level image. The applied filter of convolutional layers is 3x3 with fixed convolution stride to one. The dimension of feature maps in each convolution layer is increased gradually from 50 in layer-1 to 400 in layer-12. After three convolutional layers, a max-pooling 2×2 window with stride 2 is implemented to halve the size of feature map. At last, two fully-connected layers with 900 and 200 hidden units respectively followed by the soft-max layer to obtain the feature vector that is used for classification. We train the network using its parameters: a stochastic gradient descent 'sgdm' with momentum and mini-batch of size 100, the learning rate is 0.001. However, we use the dropout technique only in the last layer.

## 5.2 Comparison engines

In this section, we introduce the comparison engine methods used to analyze the children's handwriting quality. Indeed, the objective is to compare the 'test' samples with three correct samples known as 'models', existing in the training database, based on the complementary extracted feature vectors.

### 5.2.1 BEM based on SD-DD Method

As mentioned previously, the BEM describes a set of dynamic and geometric features designed for online handwriting trajectory modeling by breaking it into $N$ elementary strokes delimited by velocity extremums. Based on this idea, the comparison of the test sample that is represented by a set of beta stroke $T = \{T_1, T_2, .., T_N\}$, with the acquired model samples $M = \{M_1, M_2, ..., M_m\}$ is done by comparing their respective strokes using a *similarity detection* (SD) and *dissimilarity distance* (DD) measure. This method uses the algorithm 1 that makes it possible to detect for each trajectory beta stroke $T_i$ of the 'test' sample, the most similar stroke $M_j$ of the model sample indexed in a neighborhood around its current index $i$. Then, it calculates the DDs distance between these two strokes $T_i$ and $M_j$. The DD between a given test sample and the model samples noted $DD_{T_M}$ is obtained by the average of the DDs of all beta strokes composing the 'test' sample in comparison with those of the model samples and vice versa.

**Algorithm 1** SD-DD comparison method

INPUT: Sets of successive trajectory beta strokes composing successively the test sample T = $\{T_1, T_2, ..., T_n\}$, and $p$ model samples: $M_k = \{M_{1,k}, M_{2,k}, ..., M_{m,k}\}, k=1,...,p$
OUTPUT: Distance between $T$ and $M$ samples
*Initialisation:* $\boldsymbol{dist_{test-models}} = 0$;
**for all** *model sample* $M_k$ , k =1 to $p$ **do**
  **for all** *stroke test* $T_i$ , $i= 1$ **to** $n$ **do**
    **for all** *stroke model* $M_{j,k}$, $j= i-l$ **to** $i+l$ **do**
    Compute: cosine similarity ($T_i$ , $M_{i,k}$)
    **end for**
    Detection of the stroke $M_{i,k}$ most similar to $T_i$
    Compute the dissimilarity distance $\boldsymbol{DD_i}$ between $T_i$ *and* $M_{i,k}$ :
    $\boldsymbol{DD_i}$ = normalized Euclidian distance ($T_i$, $M_{i,k}$)
  **end for**
  Compute the dissimilarity distance between $T$ and $M_k$ samples
  $$\boldsymbol{dist_{T-M_k}} = \frac{\sum_{i=1}^{n} \boldsymbol{DD_i}}{n}$$
  $$\boldsymbol{dist_{test-models}} = \boldsymbol{dist_{test-models}} + \boldsymbol{dist_{T-M_k}}$$
**end for**
Compute the dissimilarity distance between $T$ and all models samples
$$\boldsymbol{DD_{T_M}} = \frac{\boldsymbol{dist_{test-models}}}{\boldsymbol{p}}$$
**end**
**Return** $\boldsymbol{DD_{T_M}}$

For each evaluation criterion, the selection of the most relevant BEM parameters is performed by using a weighting selector vector $S_p$ of the same size as the BEM vector $V_{BEM}$ by multiplying the latter by $S_p$ as described in Eq.8.

$$T_i = V_{BEMi} * S_p \quad (8)$$

For the *Shape* criterion, all the geometric parameters of BEM are considered at weight 1 while the dynamic component is attenuated by a weighting coefficient equal to 0.2 and vice versa for the *kinematic* criterion. For both *order* and *direction* criteria, the most weighted BEM parameters are the coordinates of the trajectory beta stroke endpoints.

### 5.2.2 SVM engine

SVM is considered as a powerful tool for linear and nonlinear classification based on a supervised learning algorithm. It has shown high success in many practical applications such as pattern recognition. Contrary to traditionally artificial neural networks, the basic formulation of SVM is the structural risk minimization instead of empirical risk. As shown in Fig 9.a, SVM is mostly used to determine an optimal separating hyper-plane by adopting a novel technique that maps the sample points into a high-dimensional feature space using a nonlinear transformation. It was originally designed to solve binary classification problems. However, it can be employed also to solve multiclass problems (see Fig.9.b) using several methods such as one-versus-all [29].
In our work, we have used SVM as a comparison tool that is trained with a set of features vectors from FDM and CNN respectively. The output is determined as $\hat{y}= sign\ (\langle w, T_i\rangle + b)$, where $T_i$ is the test sample feature vector, and the model is defined by $w$ called the 'decision function' and $b \in R$. The entity $\langle w, T_i\rangle + b$ can be treated as a vector of confidence score affecting the test sample to the following assessment labels (*correct shape, wrong shape, correct order, wrong order, correct direction, inversed direction*). High absolute values of one or multiple components of this vector provide the assessment labels that summarizes the evaluation report.

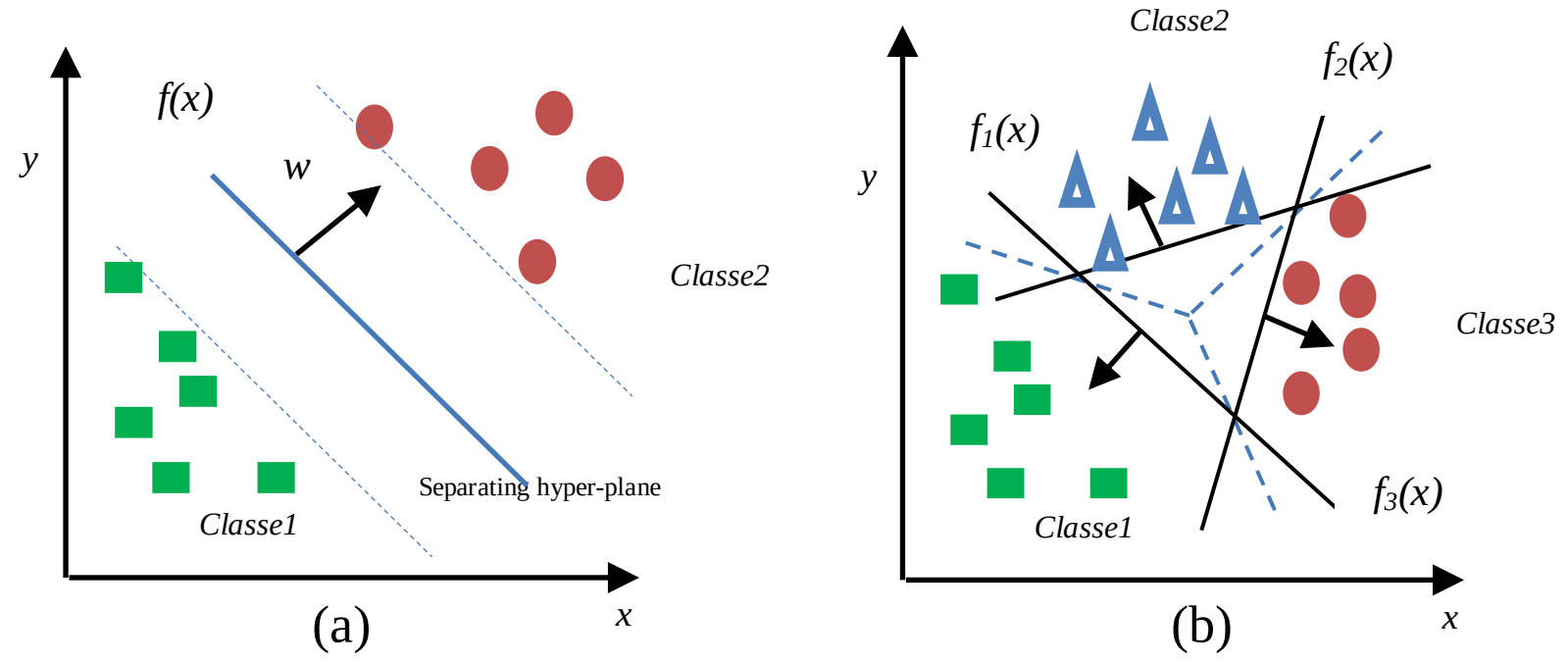

**Fig.9**. Principle of SVM. (a) two-class hyper-plane example, (b) one-versus-all method

## 6 Experimental results and discussion

This section presents experimental results that are performed on handwriting quality analysis. Datasets are firstly presented followed by the conducted experimental setup. Finally, the carried out experiments and obtained results as well as a discussion are explained.

### 6.1 Dataset

The dataset used to validate the efficiency and robustness of our system is collected by 400 children aged four to eight in different Tunisian primary school. The constructed dataset includes three subsets. Set 1 is dedicated for Arabic script that contains a block of letters (i.e.أ، ب، ع، ح، د، ر، ص، ط، س، ن، ك، ف، ي، م، ة) and words (قرأ (read), مدرسة (School), (Child) طفل, (Computer) حاسوب, (bureau) مكتب, (blackboard) سبورة, (Science) علم, (director) مدير) selected from the primary textbook. Set 2 comprises some samples of Latin characters (i.e. A, C, D, E, H, I, L, M, N, O, P, R, U). Also, we have evaluated our system with other observations from set 3 that contains some symbols such as (&, $, £, Ω). Indeed, 50 correct samples for each sequence (letters, words, symbols) are utilized for training dataset with a few samples of incorrect order and direction. However, to fully validate order and direction classifiers aims to increase the dataset, we generated additional negative samples (wrong order/direction) from positive samples by synthetically changing the order and direction. Similarly, we applied distortion technique by modifying some parameters such as the angle of inclination of the trajectory, the baselines and smoothing in order to

construct other positive/negative shapes by generating more training samples. The employed dataset is divided into a training set of 30000 samples and a test set of 10000 samples.

### 6.2 Setup

To study the impact of the already mentioned models and their combination in handwriting quality analysis, we have designed four groups of experiments. The first one is based on BEM using SD-DD comparison method. The second and third tests are realized respectively on FDM and CNN based on SVM classifier. Finally, the fourth test is founded on the fusion of the three tests.

**BEM with SD-DD**: The BEM is used to extract a set of pertinent features to characterize the five analysis criteria. The system determines for each criterion two thresholds: $T_{CC}$ and $T_{CW}$ delimiting respectively three evaluation zones: certainly correct (*CC*), Fuzzy (*F*), and certainly wrong (*CW*). As described in Eqs. (9) and (10), these thresholds are computed after statistical analysis of two distributions distances $\{DD_{C_M}\}$ and $\{DD_{W_M}\}$ separating the model samples from correct and wrong samples respectively.

$$T_{CC} = \min\left(Q_{\{DD_{C_M}\}}(u_{\max}), Q_{\{DD_{W_M}\}}(u_{\min})\right) \quad (9)$$

$$T_{CW} = \max\left(Q_{\{DD_{C_M}\}}(u_{\max}), Q_{\{DD_{W_M}\}}(u_{\min})\right) \quad (10)$$

Where $Q_{\{DD\}}(u)$ is the value of the quantile function of the $\{DD\}$ distribution at a cumulative probability of $u$ percent. $u_{max}$ and $u_{min}$ are adjustable cumulative probabilities that limit the substantial part of the distribution *DD* fixed empirically to 96% and 4% respectively.

A first normalized score $NS_1$ is assigned to the test sample relying on the correct partition using Eq. 11.

$$\begin{cases} if\ \ DD_{T_M} < T_{CC} \Rightarrow NS_1 = 1 \\ elseif\ \ DD_{T_M} > T_{CW} \Rightarrow NS_1 = 0 \\ else\ \ NS_1 = \dfrac{T_{CW} - DD_{T_M}}{T_{CW} - T_{CC}} \end{cases} \quad (11)$$

Similarly, we calculated the $DD_{T_W}$ distance separating the test sample from the set of wrong samples which represents the most common errors before converting it to a normalized score noted $NS_2$ between 0 and 1 analogously to Eq.11. As described in Eq. 12, the final score *NS* is computed as the average between $NS_1$ and the complement of $NS_2$.

$$NS = \frac{NS_1 + (1 - NS_2)}{2} \quad (12)$$

**FDM and CNN based on SVM classifier:** In this part, we have initially extracted the feature vectors using FDM and CNN models. Thereafter, these later are used as input for the multiclass SVM with RBF kernel function for making classification step. The SVM model allows characterizing such criteria: *shape, direction,* and *order* using FDM and only *shape* and *position* criteria with CNN model relying on the classifier confidence score.

**Combined models:** The fourth experiment evaluated the impact of the hybrid results of the already mentioned models. The final score assigned for each criterion is calculated as a weighted average of the scores assigned by each separate subsystem weighted by its correct classification rate (CCR).

### 6.3 Results

Our system considers three evaluation levels: a global description level which allows to classify the test script into six classes identified by a label $L_{ic}$ and a digital index $i_c$: $i_c$ = 1, $L_1$ = correct; $i_c$ = 2, $L_2$ = wrong shape; $i_c$ = 3, $L_3$ = wrong order; $i_c$ = 4, $L_4$ = wrong direction; $i_c$ = 5, $L_5$= reference line surpass; $i_c$ = 6, $L_6$ = irregular kinematics). The second level called quantitative analysis assigns a confidence score ranging from [0,100] for each criterion. The last level provides a qualitative assessment of the test script by attributing for each criterion two membership rates $R_{EB_1}$ and $R_{EB_2}$ of qualitative evaluation labels among: very well (VW), Well (W), medium (M), bad (B), very bad (VB). Thereby, the membership rates $R_{EB_1}$ and $R_{EB_2}$ are computed considering the projection of the average value of the criterion scores $S_{Crit}$ provided at the input of the decision fusion stage (see Fig.1), on the intersected linguistic fuzzy subsets among the five labels (see Fig.10).

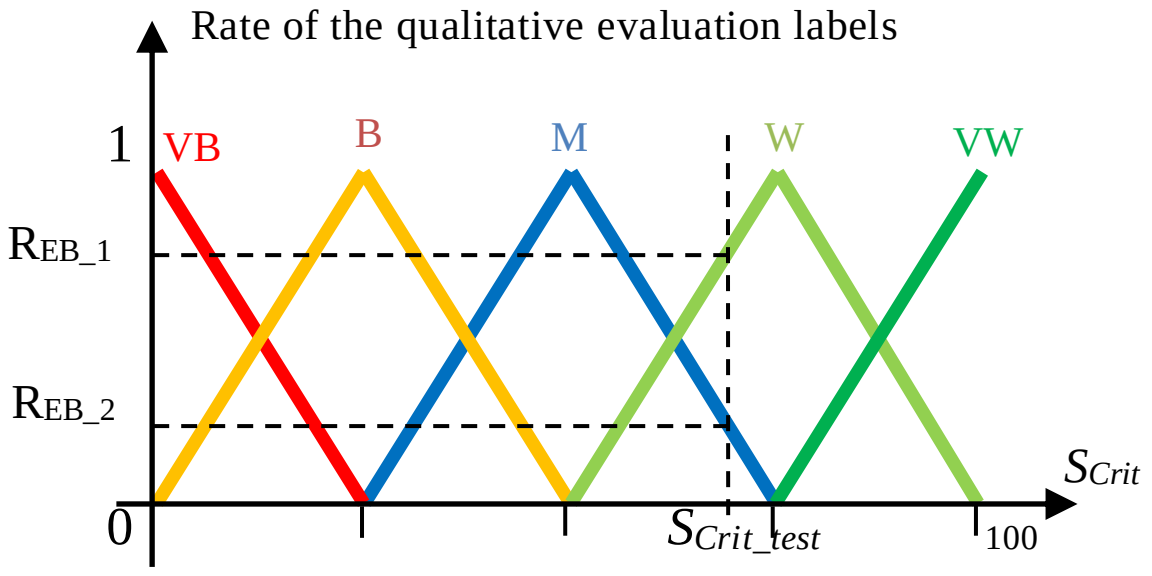

**Fig. 10.** Qualitative evaluation labels

**Table 2. Overall CCR [%] using correct/ incorrect samples of shape and direction criteria.**

| Criteria | Shape | | | | Direction | | |
|---|---|---|---|---|---|---|---|
| Models | BEM | FDM | CNN | Fusion | BEM | FDM | Fusion |
| **Arabic characters** | 96,56 | 97,16 | 98,86 | 99,02 | 98,16 | 96,26 | 98,71 |
| **Arabic words** | 94,13 | 96,50 | 97,25 | 97,96 | 97,10 | 96,13 | 98,61 |
| **Latin characters** | 96,9 | 97,02 | 97,14 | 97,92 | 99,50 | 97.11 | 99,50 |
| **Symbol** | 97,13 | 98,02 | 99,11 | 99,41 | 98.13 | 98.08 | 99.12 |

**Table 3. Overall CCR [%] using correct/ incorrect samples of order, position, and kinematic criteria.**

| Criteria | Order | | | Kinematic | Position | | |
|---|---|---|---|---|---|---|---|
| Models | BEM | FDM | Fusion | BEM | BEM | CNN | Fusion |
| **Arabic characters** | 98,75 | 97,15 | 98,75 | 98,75 | 99.00 | 97,06 | 99.00 |
| **Arabic words** | 97,62 | 95,35 | 97,84 | 97,20 | 95,77 | 93,16 | 96.11 |
| **Latin characters** | 97.30 | 96.14 | 97,72 | 98.90 | 96,51 | 95,18 | 97.13 |
| **Symbol** | 98.20 | 96.10 | 99,15 | 98,60 | 98,62 | 98,11 | 99.97 |

### 6.3.1 Global analysis

Before presenting the detail of quantitative and qualitative results of handwriting analysis, it is necessary to provide an overall legibility assessment of each criterion that involves an evaluation of the test scripts classification process applied by each model. In this process, FDM and CNN approaches operate SVM algorithms, while the global classification using BEM approach assigns the test sample to the class of index $i_c$ following the eq. 13.

$$i_c = argmax\left\{S_{shape}\ ,\ 1-S_{direction}\ ,\ 1-S_{ordre}\ ,\ 1-S_{kinematic}\ ,\ 1-S_{position}\right\} \quad (13)$$

Where: $S_{crit}$: is the score of the correspondent **crit**erion.

Tables 2, and 3 summed up the global results achieved by our system using the different feature extraction models and their fusion. We can see from these tables that our method is very effective for handwriting children's evaluation. In fact, it gives good results for characterizing the five criteria. The specific temporal information and geometric parameters provided by BEM allow us to analyze the five criteria (*shape*, *direction*, *order*, *kinematic*, and *position*). FDM is used to characterize (*shape*, *direction*, *order*) regardless of the handwriting velocity. Finally, the generic features generated by CNN allowing to apprehend only the *shape* and *position* criteria regardless of the handwriting dynamic and order aspects.

**Table 4. Quantitative results of some Arabic, Latin scripts and symbols using shape, order and direction criteria.**

| Criteria | Shape | | | Order | | Direction | |
|---|---|---|---|---|---|---|---|
| **Model** | **BEM** | **FDM** | **CNN** | **BEM** | **FDM** | **BEM** | **FDM** |
| | Arabic characters | | | | | | |
| أ | 0.95 | 0.97 | 0.99 | 0.99 | 0.95 | 1.00 | 0. 97 |
| ب | 0.94 | 0.97 | 0.98 | 0.98 | 0.96 | 0.99 | 0.95 |
| ع | 0.95 | 0.95 | 0.97 | - | - | 0.97 | 0.95 |
| ح | 0.96 | 0.96 | 0.99 | - | - | 0.99 | 0.98 |
| د | 0.98 | 0.96 | 0.99 | - | - | 1.00 | 0.97 |
| و | 0.97 | 0.97 | 0.98 | - | - | 1.00 | 0.99 |
| ر | 0.98 | 0.97 | 0.99 | - | - | 1.00 | 0.97 |
| ص | 0.94 | 0.96 | 0.97 | - | - | 0.97 | 0.93 |
| ط | 0.95 | 0.95 | 0.95 | 0.97 | 0.93 | 0.96 | 0.95 |
| س | 0.90 | 0.92 | 0.93 | - | - | 0.93 | 0.90 |
| ن | 0.93 | 0.96 | 0.98 | 0.95 | 0.92 | 0.95 | 0.95 |
| ك | 0.94 | 0.95 | 0.99 | 1.00 | 1.00 | 1.00 | 0.96 |
| ف | 0.91 | 0.91 | 0.92 | 0.99 | 0.96 | 0.97 | 0.97 |
| ي | 0.88 | 0.94 | 0.96 | 0.96 | 0.94 | 0.96 | 0.97 |
| م | 0.90 | 0.91 | 0.95 | - | - | 1.00 | 1.00 |
| ة | 0.93 | 0.96 | 0.96 | 0.99 | 0.98 | 1.00 | 0.97 |
| | Arabic words | | | | | | |
| قرأ | 0.89 | 0.92 | 0.95 | 1.00 | 0.93 | 1.00 | 0.95 |
| مدرسة | 0.92 | 0.93 | 0.96 | 0.97 | 0.94 | 0.98 | 0.92 |
| طفل | 0.94 | 0.96 | 0.97 | 0.96 | 0.93 | 0.99 | 0.91 |
| | Latin characters and symbols | | | | | | |
| A | 0.94 | 0.96 | 0.98 | 0.98 | 0.96 | 0.99 | 0.97 |
| C | 0.97 | 0.97 | 0.99 | - | - | 1.00 | 0.99 |
| D | 0.93 | 0.96 | 0.98 | 0.99 | 0.97 | 1.00 | 0.95 |
| E | 0.95 | 0.94 | 0.97 | 1.00 | 0.95 | 0.99 | 0.97 |
| H | 0.96 | 0.96 | 0.98 | 0.99 | 0.97 | 1.00 | 0.97 |
| $ | 0.98 | 0.98 | 1.00 | 0.99 | 0.97 | 1.00 | 0.98 |
| & | 0.96 | 0.96 | 0.98 | - | - | 1.00 | 1.00 |

**Table 5. Quantitative results of some test samples of Arabic, Latin scripts and symbols using position, and kinematic criteria.**

| Criteria | Position of the trace | | Kinematic |
|---|---|---|---|
| **Model** | **BEM** | **CNN** | **BEM** |
| أ | 0.95 | 0.97 | 0.99 |
| ب | 0.94 | 0.91 | 0.98 |
| ع | 0.95 | 0.95 | 0.99 |
| ح | 0.96 | 0.91 | 0.92 |
| د | 0.98 | 0.96 | 0.97 |
| و | 0.97 | 0.97 | 0.98 |
| ر | 0.98 | 0.97 | 0.98 |
| ص | 0.94 | 0.96 | 0.91 |
| ط | 0.95 | 0.95 | 0.88 |
| س | 0.93 | 0.91 | 0.87 |
| ن | 0.93 | 0.96 | 0.95 |
| ك | 0.94 | 0.95 | 0.83 |
| ف | 0.91 | 0.91 | 0.96 |
| ي | 0.93 | 0.95 | 0.96 |
| م | 0.90 | 0.91 | 0.89 |
| ة | 0.93 | 0.96 | 0.99 |
| قرأ | 0.96 | 0.98 | 1.00 |
| مدرسة | 0.92 | 0.93 | 0.97 |
| طفل | 0.94 | 0.96 | 0.96 |
| A | 0.94 | 0.96 | 0.98 |
| C | 0.97 | 0.92 | 0.97 |
| D | 0.96 | 0.98 | 0.99 |
| E | 0.95 | 0.94 | 1.00 |
| H | 0.96 | 0.96 | 0.99 |
| $ | 0.98 | 0.98 | 0.99 |
| & | 0.96 | 0.96 | 1.00 |

### 6.3.2 Quantitative analysis

The accuracy $AS_{accuracy}$ of the quantitative evaluation obtained by the different individual approaches implemented in the proposed system is described in Eq.14. It is computed by comparing the provided scores $AS_{predicted}$ at the output of their respective modules with the scores $AS_{expected}$ assigned by the teachers during the database collection.

$$AS_{accuracy} = 1 - \left| AS_{predicted} - AS_{expected} \right| \quad (14)$$

We evaluate the performance of our handwriting analysis system for all supported sequences (Arabic letters and words, Latin letters, and symbols) on the collected test dataset. The obtained quantitative results of some observations are shown in Table 4 and 5. It shows that the best confidence score of *shape* criterion is achieved by using CNN and FDM rather than BEM (see Table 4). This can be explained by the use of visual and curvature generic features generated by CNN and FDM respectively. Also, the BEM approach is more accurate for the analysis of *order* and *direction* criteria than a network of score regression or prediction like the association FDM-SVM. This robustness is due to the precise features to characterize the change of direction and order of elementary strokes. We notice also that the *kinematic* criterion allows us to distinguish the nature of writing (fast or slow) based on a powerful BEM model (see Table 5). This is due to the use of dynamic parameters that give the overall temporal properties of the neuromuscular networks implicated in motion generation, and the geometric properties of all the muscles and joints inducted to execute the movement. Likewise, we can see from this table that the obtained results using BEM and CNN for *position of the trace* criterion are very powerful which aid children during learning process.

### 6.3.3 Qualitative analysis

As illustrated in Fig. 11, Fig. 12, and Fig. 13, qualitative assessments of the used criteria are presented respectively for some samples of Arabic, Latin scripts, and symbols. As shown in Fig.11 (a), the test samples, their number, and their global confidence scores are displayed on the top left corner, right, and left bottom corners respectively. Five levels of qualitative assessment are considered for the confidence score: VW (green), W (dark green), M (blue), B (orange), and VB (red) corresponding to a uniform partition of confidence ranges.

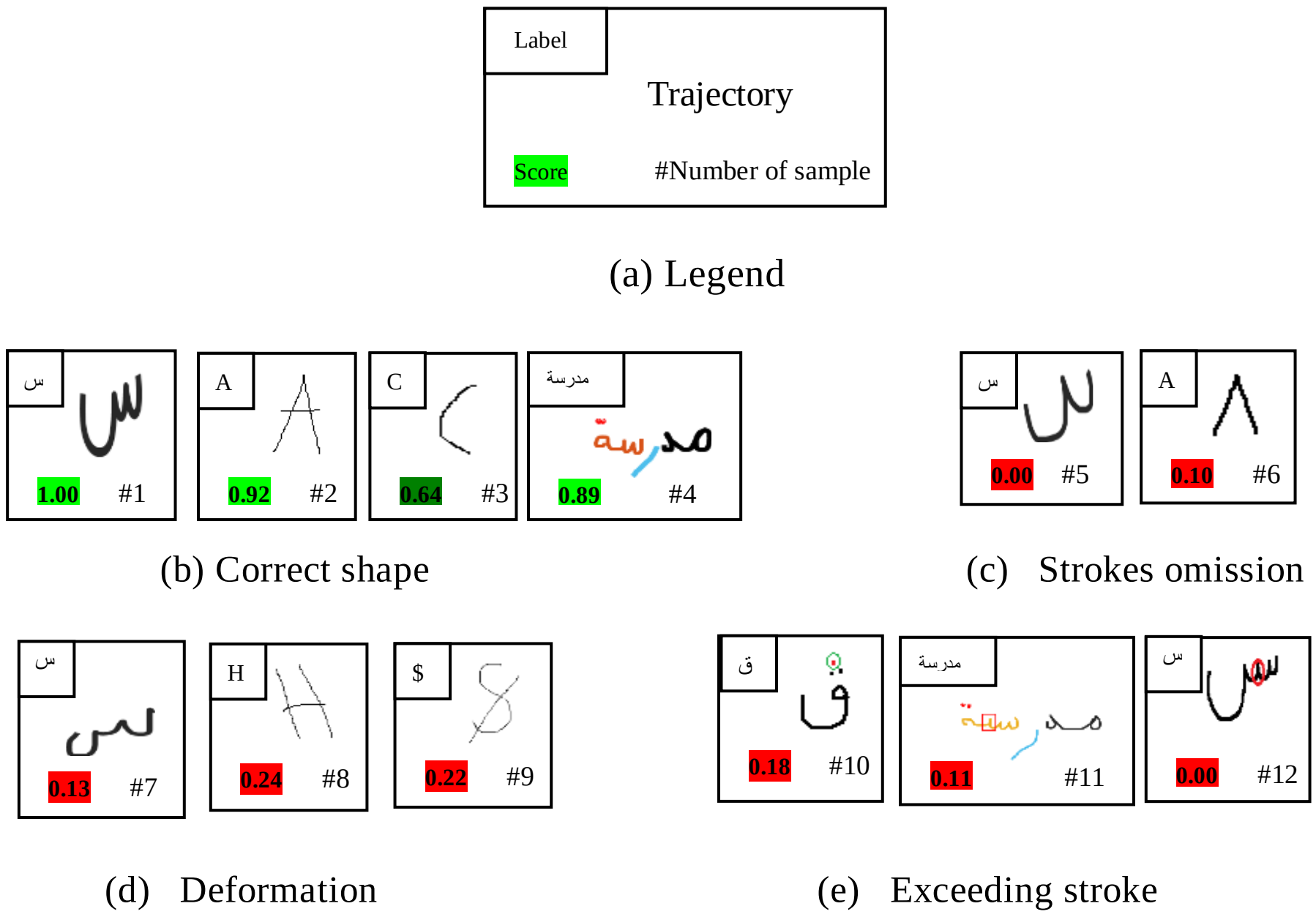

**Fig. 11.** Qualitative evaluation labels on the shape criterion with legend (a), correct shape (b), wrong shape due to omission strokes (c), deformation stroke (d) and exceeding stroke (e).

Results show that the proposed method is rather generic which gives pertinent analysis scores for both cursive and non-cursive children's handwriting. As depicted in Fig 11, it is able to determine the correct shape as shown in samples #1 to #4 of Fig.11. (b), detect the wrong shapes caused by strokes omission (#5 and #6 of Fig.11.c), strokes deformation (#7, #8, and #9 of Fig.11.d), or exceeding strokes (The red bounding box in observation #10, #11, and #12 of Fig.11.d).

Also, qualitative results demonstrate the ability of the system to analyze correctly the direction and order criteria. Fig.12.a) and Fig.12.b) show the evaluation of various numbers of samples according to these criteria. We notice that the mentioned rate represents the average score attributed to the analyzed sample according to the considered criteria which result from the fusion of the used models. And the color of its highlight represents the qualitative interpretation.

Furthermore, the combination method allows us to analyze the handwriting trajectory kinematics and the respect of the handwriting reference lines. Fig 13.a) shows the difference between the BEM dynamic profiles modeling and their results of samples' analysis that are written quickly (#19, # 21) to others carefully (#20). Similarly, the cases of the trace position errors corresponding to the overflow of the reference line have been penalized comparing to those traced with respect to the baseline in terms of criterion score and its qualitative interpretation (color of score highlight, see Fig.13 b).

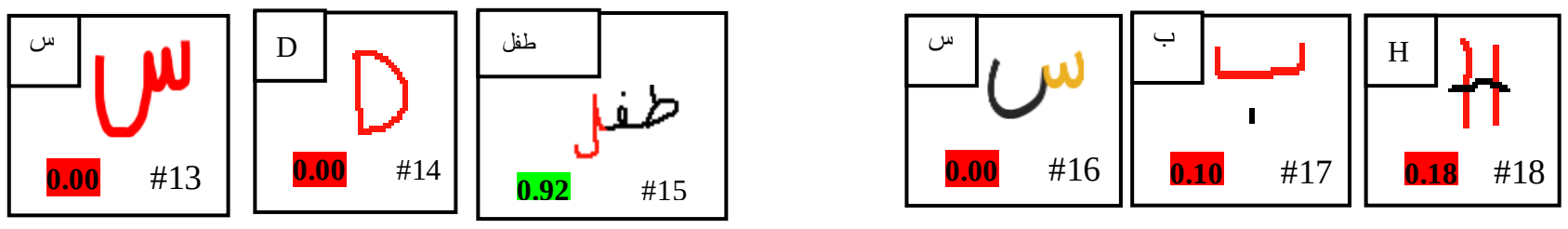

(a) Direction (b) Order

**Fig. 12.** Qualitative results on both direction and order criteria.

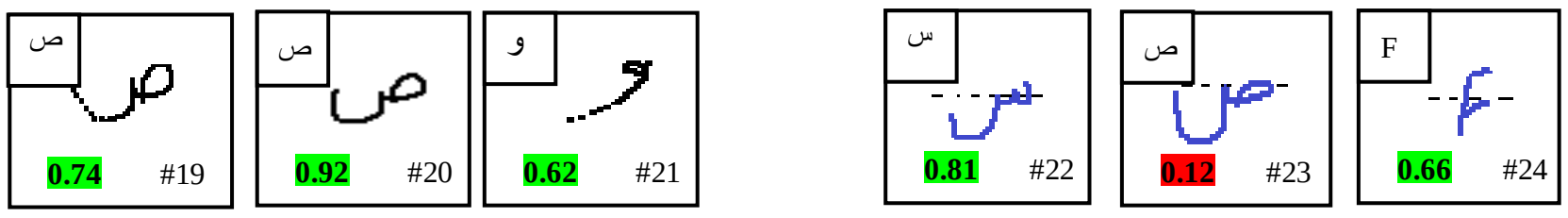

(a) Kinematic (b) Position of the trace

**Fig. 13.** Qualitative results on Kinematic and position criteria.

## 6.4 Discussion

We discussed below the strengths and limitations of the present framework as well as the comparative performance analysis with other existing online handwriting analysis systems.

### 6.4.1 Strengths of the present framework

Our experimental results show that the proposed analysis criteria and adopting extractor models provide superior performance over existing online Latin handwriting analysis systems. The major strengths of our framework are as following:

- The ability to analyze both cursive and non-cursive multilingual scripts based on the complementarity of feature extraction models.

- The present framework can evaluate the children's handwriting quality with different writing styles such as presented in Fig.14.a) and Fig.14.b) for correct shape of Arabic Letters 'ع' and 'ح', Fig.14.c) and Fig.14.d) of Latin letters '*a*' and '*i*' respectively.

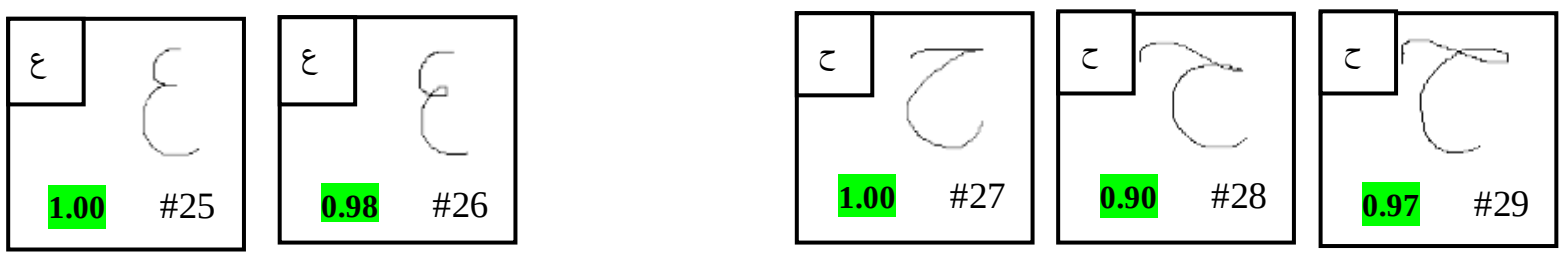

(a) Arabic Letter Ain 'ع' (b) Arabic Letter Haa 'ح'

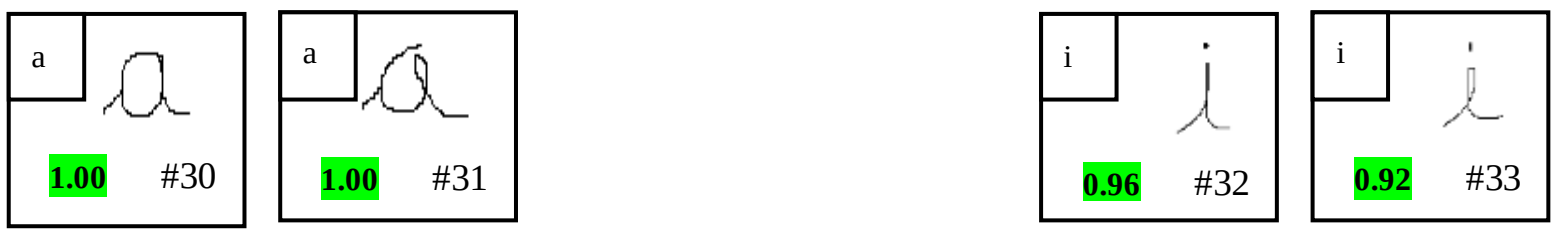

(c) Latin letter 'a' (d) Latin letter 'i'

**Fig. 14.** Handwriting evaluation quality with different writing styles.

- Our proposal is original compared to other existent researches, since it is based on the combination of two global evaluation approaches: the first one employs SD-DD comparison engine based on BEM with three model samples and the six mentioned class. The second approach employs the hybrid of FDM and deep learning CNN features with SVM engine.

### 6.4.2 Limitation

While analyzing the limitations of the proposed framework in handwriting quality analysis system, it has been noted that most of the errors generated in the two script language Arabe and Latin are mainly due to a poor appreciation of the *shape* criterion using BEM approach when the user writes very quickly or slowly. In fact, in these two cases, the BEM generates an irregular number of strokes very different from the model samples, which affects the geometry of the strokes and thus the score of the shape criterion. On the other hand, the two modules FDM and CNN compensate for this drawback at the level of the fusion step. However, the BEM approach is a more appropriate module for the dynamic criterion evaluation.

### 6.4.3 Comparative performance analysis

As already mentioned, there are no available researches for the analysis of online Arabic handwriting script as there are very few significant studies on Latin script in the literature. All the existing analytical works in Latin script have used different datasets that are not always available, so these studies are not comparable directly. This did not prevent us to compare the performance of the proposed framework with these existing studies [1], [2], [15] to get an idea of comparative performance analysis.

We retain that the performance of our system in Latin script is comparable to that of the reference commercial systems [2], and [15], in the *order* criterion with 97.62% of CCR and a slightly advantageous in terms of correct analysis for *shape* and *direction* criteria with 96.6% and 98.3% of CRR respectively. On the other hand, our system is distinguished by the possibility of analyzing other essential criteria such as *kinematic* (dynamics) and *respect of reference lines* which are two important aspects in the learning step.

## 7 Conclusion and Future Work

This paper introduces a new analysis method for children handwriting quality allowing them to deal with both cursive and not cursive handwriting multi-script. The key emphasis of our project is to analyze the handwriting quality of children's school and provide them with relevant real-time feedback based on multiple criteria. The proposed system uses the combination of three models for handwriting representation. The BEM offers the possibility of recovering dynamic and geometric aspects in online handwriting trajectory modeling. It is characterized by the strong ability to analyze five main criteria (*shape, direction, order, kinematic, position*). Also, FDM offers a specific feature allowing to analyze precisely *shape, order,* and *direction* criteria. Finally, we employed CNN, a visual abstraction model, to describe the final *shape* of the post-drawing view.

The evaluation results and the computed scores assigned for each criterion are conducted using two different methods: SD-DD and SVM-based comparison engines. The efficiency of the single corresponding models, and the combination of them, was evaluated in experiments using different observation of the broad database. By combining models, the obtained results are promising and suggest that our proposed method is well suited to handwriting quality analysis. We have also observed that the used extracted features are rather generic and their applicability in other scripts as Persian, Urdu, English, etc., is interesting, so we envisage it as future work. Finally, we maintain that our proposal can be appropriated for other applications such as the pre-diagnosis of neuromuscular pathologies and the difficulties of dyslexia.

## Acknowledgements

The research leading to these results has received funding from the Ministry of Higher Education and Scientific Research of Tunisia under the grant agreement number LR11ES4.